\documentclass[graybox]{svmult}

\usepackage{mathptmx}       
\usepackage{helvet}         
\usepackage{courier}        
\usepackage{makeidx}         
\usepackage{graphicx}        
\usepackage{multicol}        
\usepackage[bottom]{footmisc}
\usepackage{url}
\usepackage{amsmath, amssymb, graphics}

\newcommand{\mathsym}[1]{{}}
\newcommand{\unicode}[1]{{}}

\makeindex             

\begin{document}
\title*{Path following and numerical continuation methods for non-linear MEMS and NEMS}

\author{Peter G. Steeneken and Jiri Stulemeijer}

\institute{Peter G. Steeneken \at NXP-TSMC Research Center, NXP Semiconductors, HTC 4, 5656 AE Eindhoven, the Netherlands, \email{peter.steeneken@nxp.com}
\and Jiri Stulemeijer \at Epcos Netherlands, 6546 AS Nijmegen, The Netherlands \email{jiri.stulemeijer@epcos.com}}

\maketitle

\abstract{Non-linearities play an important role in micro- and nano- electromechanical system (MEMS and NEMS) design. In common electrostatic and magnetic actuators, the forces and voltages can depend in a non-linear way on position, charge, current and magnetic flux. Mechanical spring structures can cause additional non-linearities via material, geometrical and contact effects. 
For the design and operation of non-linear MEMS devices it is essential to be able to model and simulate such non-linearities. However, when there are many degrees of freedom, it becomes difficult to find all equilibrium solutions of the non-linear equations and to determine their stability. In these cases path following methods can be a powerful mathematical tool.
In this paper we will show how path following methods can be used to determine the equilibria and stability of electromechanical devices. Based on the energy, work and the Hamiltonian of electromechanical systems (section \ref{enwork}), the equations of motion (section \ref{eqmot}), the equilibrium (section \ref{equilibrium}) and stability conditions (section \ref{stability}) are derived. Examples of path following simulations (section \ref{pathfollowing}) in Mathematica (section \ref{switchexample}), \textsc{Matcont} (section \ref{matcontex}) and using FEM methods in \textsc{Comsol} (section \ref{fem}) are given to illustrate the methods.}

\pagebreak

\section{Energy and work in electromechanical systems}
\label{enwork}

MEMS and NEMS devices can often be treated using classical physics. Interactions between the electric and magnetic fields can usually be neglected in MEMS and NEMS devices, because the system dimensions are small compared to the electromagnetic wavelength and are large enough to neglect quantum effects. In these approximations the total internal energy of the system $U_{tot}=\sum U$ and total work done by external sources $W_{tot}=\sum W$ can be expressed as a sum over different types of energy and work, given by equations (\ref{ukin}-\ref{wstrain}):

\begin{eqnarray}
U_{kin}&=&\int_{\mathcal{V}} {\frac{1}{2} \frac{\vec{p}_m^2}{\rho_m} {\rm d}\mathcal{V}}=\frac{1}{2} \sum_i {\frac{p_{m,i}^2}{m_i}} \label{ukin} \\
U_{grav}&=&\int_{\mathcal{V}} {\frac{1}{2} V_G \rho_e {\rm d}\mathcal{V}} =  \frac{1}{2} \sum_{i} \left( m_i \sum_{j}  \frac{G}{r_{ij}} m_j  \right) \label{ugrav} \\
U_{mag}&=& \int_{\mathcal{V}} {\frac{1}{2} \vec{J} \cdot \vec{A} {\rm d}\mathcal{V}} =\frac{1}{2} \sum_{i} \left( \Phi_i \sum_{j}  L^{-1}_{ij} \Phi_j  \right) \label{umag} \\
U_{el}&=&\int_{\mathcal{V}} {\frac{1}{2} V \rho_e {\rm d}\mathcal{V}} = \frac{1}{2} \sum_{i} \left( Q_i \sum_{j}  C^{-1}_{ij} Q_j  \right) \label{uel} \\
U_{strain}&=&\int_{\mathcal{V}} {\frac{1}{2} \vec{\epsilon} \cdot \vec{\sigma} {\rm d}\mathcal{V}}=\frac{1}{2}\sum_{i} {\left(  x_i \sum_{j} {k_{ij}  x_j}\right)} \label{ustrain} \\
W_{kin} &=& \int_{\mathcal{V}} {\int_0^{\vec{p}_m} {\vec{v}_{ext} \cdot \vec{dp}_m} {\rm d}\mathcal{V}} =\sum_i {\int_0^{p_{m,i}} {v_{ext,i} {\rm d}p_{m,i}}} \label{wkin}\\
W_{grav} &=& \int_{\mathcal{V}} { \int_0^{\rho_m} {V_{G,ext} {\rm d}\rho_m} {\rm d}\mathcal{V}}=\sum_i {\int_0^{x_i} {m_i g_{ext,i} {\rm d}x_i}} \label{wgrav}\\
W_{mag} &=& \int_{\mathcal{V}} {\int_0^{\vec{A}}{\vec{J}_{ext} \cdot \vec{dA}} {\rm d}\mathcal{V}}=\sum_i {\int_0^{\Phi_i} {I_{ext,i} {\rm d}\Phi_i}} \label{wmag}\\
\label{wel} W_{el} &=& \int_{\mathcal{V}} {\int_0^{\rho_e}{V_{ext} {\rm d}\rho_e} {\rm d}\mathcal{V}}=\sum_i {\int_0^{Q_i} {V_{ext,i} {\rm d}Q_i}} \\
W_{strain} &=& \int_{\mathcal{V}} {\int_0^{\vec{\epsilon}} {\vec{\sigma}_{ext} \cdot \vec{d\epsilon}} {\rm d}\mathcal{V}} =\sum_i {\int_0^{x_i} {F_{ext,i} {\rm d}x_i}} \label{wstrain} 
\end{eqnarray}

These expressions are given to show that there is a striking similarity between energy and work equations in different physical domains. As a result of this, their physics can be analyzed by the same mathematical methods. Moreover, physical coupling and transduction between domains, which are often encountered in electromechanical systems, can also be analyzed. In this paper these mathematical methods, based on the conservation of energy ($U_{tot}=W_{tot}$), will be discussed and some examples will be given of how numerical path following can be used to analyze such systems. 

The energy and work expressions in each of the equations can be written in terms of volume $\mathcal{V}$ integrals over energy densities, which can be treated numerically by finite element methods. The work equations also include an inner integral, which expresses the fact that the work done can be path dependent. If the system consists of discrete elements, simplified energy expressions on the right side of (\ref{ukin}-\ref{wstrain}) can be given, which can substantially reduce the number of degrees of freedom and are more suitable to treat systems of particles or electrical networks with lumped element components. 

The variables in equations (\ref{ukin}-\ref{wstrain}) are given as follows. $\rho_m$ is the mass density, $\vec{p}_m$ is the mass momentum density. $V_G$ is the gravitational potential and $G$ is the gravitational constant. $V$ is the electric potential, $\rho_e$ is the charge density, $\vec{J}$ is the total current density and $\vec{A}$ is the magnetic vector potential. $\vec{\sigma}$ and $\vec{\epsilon}$ are the mechanical stress and strain. The expressions on the right side of the equations are written as sums over discrete charges $Q_i$, magnetic fluxes $\Phi_i$ and masses $m_i$ at positions $x_i$ and with mass momentum $p_{m,i}$, which are coupled via effective springs $k_{ij}$, gravitational potential $G/r_{ij}$, capacitors $C_{ij}$ and inductors $L_{ij}$. The components of the inverse capacitance and inductance matrices $\vec{C}^{-1}$ and $\vec{L}^{-1}$ relate potential to charge $V_j= \sum_i C^{-1}_{ij} Q_i$ and relate current to magnetic flux $I_j=\sum_i L^{-1}_{ij} \Phi_i$. The total work $W_{tot}$ supplied to the system is a sum of the work from external gravitational $V_{G,ext}$, current density $\vec{J}_{ext}$, electric $V_{ext}$, magnetic $\vec{A}_{ext}$ and stress $\vec{\sigma}_{ext}$ fields. The kinetic work term $W_{kin}$ is needed to account for a relative speed difference $\vec{v}_{ext}$ between the frame of reference of the system and the frame of the observer. The work can also be performed on discrete elements by external gravitational acceleration $g_{ext,i}$, current $I_{ext,i}$ or voltage $V_{ext,i}$ sources or external (stress) forces $F_{ext,i}$. For more details on equations (\ref{ukin}-\ref{wstrain}) the reader can refer to standard texts on classical mechanics and electromagnetics. 

External thermal sources at temperature $T_{ext}$, which add heat $Q_h$ to the system, can be treated by similar methods, although this requires keeping track of the entropy $S$ and temperature $T$ in the system. According to the first law of thermodynamics (${\rm d} U_{tot}= \delta Q_h + \delta W_{tot}$), this requires adding an integral $Q_{h}= \int_0^{S}{T_{ext} {\rm d}S}$ to the work equations. The internal thermal energy of the material is in essence a sum of the kinetic energy of the atoms which can be expressed by a thermal energy term $U_{therm,kin}=\int_{\mathcal{V}} \int_0^T c(T) \rho_m {\rm d}T {\rm d}\mathcal{V}$, where $c(T)$ is the specific heat capacity.

\begin{figure}[]
\includegraphics[scale=0.9]{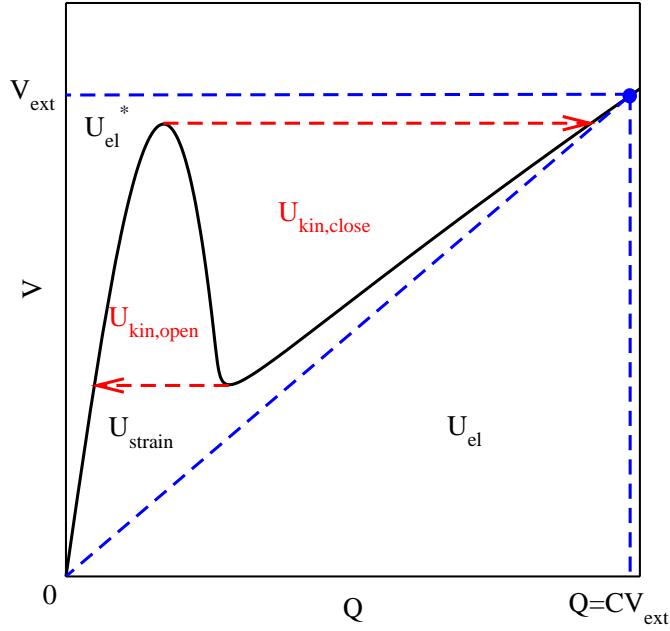}
\caption{Energy and coenergy in a voltage controlled capacitive MEMS switch. The areas between the lines in this graph represent the internal electric energy $U_{el}$ and strain energy $U_{strain}$ (between blue dashed and black solid lines), the kinetic energy $U_{kin}$ (between red dashed arrows and black solid lines) and the path-dependent electric coenergy $U^*_{el}$ of a capacitive MEMS switch. The black curve will be calculated in section \ref{switchexample} and represents the static equilibria of a capacitive MEMS switch which is actuated by a voltage $V_{ext}$. If the device follows the black path, no kinetic energy is generated because the device is always in static equilibrium. The red dashed arrows represent the hysteresis path that is followed when the switch is actuated using voltage control. In this case the areas between the red and black lines is the generated kinetic energy during pull-in ($U_{kin,close}$) and pull-out ($U_{kin,open}$). The sum of both energies $U_{kin,close}+U_{kin,open}$ is the dissipated energy during one switching cycle.}
\label{Energy}       
\end{figure}

The work provided by an electrical source in equation (\ref{wel}) can be rewritten using integration by parts: 
$W_{el}=\int_{\mathcal{V}} {(V_{ext} \rho_e - \int_0^{V_{ext}} {\rho_e {\rm d}V_{ext}}) {\rm d}\mathcal{V}}$. It is often convenient to define the coenergy or complementary energy as the Legendre transform of the work such that 
$U^*_{el}\equiv \int_{\mathcal{V}} {V_{ext} \rho_e {\rm d}\mathcal{V}}-W_{el}
=\int_{\mathcal{V}} {\int_0^{V_{ext}}{\rho_e {\rm d}V_{ext}} {\rm d}\mathcal{V}}$. Figure \ref{Energy} shows how energies and coenergies can be represented graphically for a voltage controlled capacitive MEMS switch example which is treated in section \ref{switchexample} and figure \ref{Examplecalc}.

\section{Equations of motion and energy conservation}
\label{eqmot}

The Hamiltonian $\mathcal{H}$ is the difference between the total internal energy $U_{tot}$ and the total supplied work $W_{tot}$. It can be described by a set of generalized coordinates $q_i$ with $(i=1\ldots N)$ and corresponding generalized momenta $p_i$: 
\begin{equation}
\mathcal{H}(q_1,\ldots ,q_N,p_1,\ldots ,p_N)=U_{tot}-W_{tot}
\label{H}
\end{equation}
From the Hamilton equations the generalized forces $F_i$ and speeds $v_i$ are given by:
\begin{equation}
F_i=\dot{p}_i= \frac{-\partial \mathcal{H}}{\partial q_i} \; , \; \; \; \; v_i=\dot{q}_i=\frac{\partial \mathcal{H}}{\partial p_i}
\label{Hamilton} 
\end{equation}

The total internal energy $U_{tot}$ in the system equals the total work $W_{tot}$ done on the system and therefore the Hamiltonian $\mathcal{H}$, which is the difference between internal energy and work $\mathcal{H}=U_{tot}-W_{tot}$, is conserved. This follows from the Hamilton equations (\ref{Hamilton}) because:
\begin{equation}
\frac{{\rm d} \mathcal{H}}{{\rm d} t}= \sum_i \left( \frac{\partial \mathcal{H}}{\partial q_i} \dot{q}_i+\frac{\partial \mathcal{H}}{\partial p_i} \dot{p}_i \right)=0
\label{Hamiltoncons} 
\end{equation}

Let us choose as generalized mechanical coordinates the position $q_{mech,i}=x_i$ with as corresponding generalized momentum $p_{mech,i}=p_{m,i}$, and for the generalized electromagnetic coordinates the charge $q_{em,i}=Q_i$ with the magnetic flux $p_{em,i}=\Phi_i$ as corresponding generalized momentum. 

If the parameters $k_{ij}$, $L^{-1}_{ij}$ and $C^{-1}_{ij}$ are constants, applying the Hamilton equations (\ref{Hamilton}) to equations (\ref{ukin}-\ref{H}) gives the equations of motion:
\begin{eqnarray}
F_{mech,i}&=&\dot{p}_{m,i}=-\sum_{j} {k_{ij}  x_j}+m_i g_{ext,i}+F_{ext,i}  \; ,  \; \; \; v_{mech,i}=\dot{x}_i = p_{m,i}/m_i \label{eqmotmech}\\
F_{em,i}&=&\dot{\Phi}_i=-\sum_{j} C^{-1}_{ij} Q_j + V_{ext,i}  \; , \; v_{em,i}=\dot{Q}_i = \sum_{j} L^{-1}_{ij} \Phi_j +I_{ext,i} \label{eqmotem}
\end{eqnarray}
This gives us Newton's second law, Hooke's law and Faraday's law of electromagnetic induction. 

However, if for example the spring constants, inductance and capacitance parameters $k_{ij}$, $L^{-1}_{ij}$ and $C^{-1}_{ij}$ depend on the mechanical positions $x_i$, which is often the case in electromechanical systems, the mechanical and electromagnetical problems become coupled and the mechanical forces become:
\begin{eqnarray}
\label{Fmech}
F_{mech,i}&=&-\sum_{j} {k_{ij}  x_j}-\frac{1}{2} \sum_{i,j} {\frac{\partial k_{ij}}{\partial x_i} x_i x_j}-\frac{1}{2} \sum_{i,j} {\frac{\partial C^{-1}_{ij}}{\partial x_i} Q_i Q_j}-\frac{1}{2} \sum_{i,j} {\frac{\partial L^{-1}_{ij}}{\partial x_i} \Phi_i \Phi_j} \nonumber \\
& & + m_i g_{ext,i}+F_{ext,i}
\end{eqnarray}
This shows that if the parameters depend on $x_i$ the generalized mechanical force $F_{mech,i}$ becomes the sum of coupling forces from different physical domains, like the spring force, electrostatic and magnetostatic forces. In equilibrium this sum of coupling forces is zero.

\section{Electromechanical equilibrium}
\label{equilibrium}

The system is defined to be in electromechanical equilibrium if all generalized forces $F_i$ are zero. In equilibrium, the gradient of the Hamiltonian is therefore also zero as shown by equations (\ref{eqeq},\ref{eqF}):
\begin{eqnarray}
\label{eqeq}
F_i = \dot{p}_i&=& \frac{-\partial \mathcal{H}}{\partial q_i}=0 \\
\label{eqF}
\vec{F}=-\nabla_{\vec{q}} \mathcal{H} &=& - \left( \frac{\partial \mathcal{H}}{\partial q_1}, \ldots , \frac{\partial \mathcal{H}}{\partial q_N} \right) = 0
\end{eqnarray}

For a specific set of generalized momenta $p_i$ and generalized external parameters $\mathcal{F}_{gen}=\{ m g_{ext}, F_{ext}, V_{ext}, I_{ext} \}$, equation (\ref{eqeq}) gives $N$ generalized force equations with $N$ unknowns which can be solved to obtain the equilibrium state of the system. If the energy function has only one variable parameter $\mathcal{F}$ and all other parameters are fixed, the equilibrium state $\vec{q}_{eq,0}$ at $\mathcal{F}=\mathcal{F}_0$ thus satisfies:
\begin{equation}
\label{Feq}
\vec{F} (\vec{q}_{eq,0}, \mathcal{F}_0 ) = -\nabla_{\vec{q}} \mathcal{H} (\vec{q}_{eq,0}, \mathcal{F}_0)=0
\end{equation}

Often equilibria exist for which all generalized speeds $v_i$ are zero. These equilibria are called {\it static} equilibria or {\it stationary} states. The equilibrium solutions of a non-linear system can be found using a powerful mathematical technique, which is called path following or numerical continuation \cite{Allgower,Doedel}.

\section{Path following and numerical continuation methods}
\label{pathfollowing}

Finding all equilibrium states of an electromechanical system is a difficult problem that cannot be solved in general because it requires one to check the whole $N$ dimensional coordinate space for solutions of equation (\ref{eqeq}), for each value of the external parameters $\mathcal{F}_{gen}$ and momenta $p_i$. However, when an equilibrium solution of the equation (\ref{eqeq}) is known for a certain set of parameters $p_i$, $\mathcal{F}_{gen}$, it is possible to efficiently find additional solutions using numerical path following techniques \cite{Allgower,Doedel}.

Recently it has been shown that path following techniques can be very useful for the analysis of electrostatic MEMS devices with non-linear contact forces \cite{Stulemeijer}. The technique can in fact be applied to almost any non-linear electromechanical system described by equations (\ref{ukin}-\ref{wstrain}). In MEMS and NEMS devices, the system is usually controlled by a single external parameter $\mathcal{F}$, which can for example be the voltage or current from an external source for electromechanical actuators, but can also be an acceleration or gas pressure force in sensor systems. The 1-dimensional external parameter $\mathcal{F}$ with the $N$-dimensional coordinate space $\vec{q}$ form a $(N+1)$-dimensional space. 

The implicit function theorem implies \cite{Allgower,Doedel} that {\it if} one equilibrium solution $\vec{q}_{eq,0}$ with $\vec{F}(\vec{q}_{eq,0}, \mathcal{F}_0)=0$ exists at external parameter value $\mathcal{F}_0$, {\it then} it has to be part of a continuous 1-dimensional equilibrium solution curve $\vec{q}_{eq}(\mathcal{F})$ at parameter values $\mathcal{F}$ in $(N+1)$-dimensional space with $\vec{q}_{eq,0}=\vec{q}_{eq}(\mathcal{F}_0)$. It is therefore often possible to start at a known equilibrium solution $\vec{q}_{eq}(\mathcal{F}_0)$ and determine the other solutions by following the curve ($\vec{q}_{eq}(\mathcal{F}),\mathcal{F})$.

Path following methods usually employ a predictor-corrector algorithm \cite{Allgower} to follow an equilibrium curve. In the predictor step, a new solution is predicted, usually by taking a step in direction of the tangent of the solution curve. In the corrector step, the predicted point is brought back on the solution curve, usually by an iterative method.   
For the predictor step it is necessary to determine the tangent direction to the solution curve. Let us assume that there exists an infinitesimal vector tangent to the solution curve $\vec{dq}_{eq,0}({\rm d}\mathcal{F}_0)$, with a corresponding infinitesimal parameter change ${\rm d}\mathcal{F}_0$ such that there is another equilibrium state at $(\vec{q}_{eq}(\mathcal{F}), \mathcal{F})=(\vec{q}_{eq,0}+\vec{dq}_{eq,0}, \mathcal{F}_0+{\rm d}\mathcal{F}_0)$. Because the new state should also satisfy equation (\ref{eqF}) the predictor algorithm can determine the tangent direction vector $(\vec{dq}_{eq,0},{\rm d}\mathcal{F}_0)$ in $N+1$ dimensional space by solving the $N$ equations:
\begin{eqnarray}
 \vec{F}(\vec{q}_{eq,0}+\vec{dq}_{eq,0}, \mathcal{F}_0+{\rm d}\mathcal{F}_0)&-&\vec{F} (\vec{q}_{eq,0}, \mathcal{F}_0) \nonumber \\ 
\label{Hessian}
= \vec{J}_{\vec{F}(\vec{q}_{eq,0}, \mathcal{F}_0)} \vec{dq}_{eq,0} &=& -\vec{H}_{\mathcal{H} (\vec{q}_{eq,0}, \mathcal{F}_0)} \vec{dq}_{eq,0} = 0
\end{eqnarray}
Where $\vec{J}_{\vec{F}(\vec{q}_{eq,0}, \mathcal{F}_0)}$ is the Jacobian of $\vec{F}$ at $(\vec{q}_{eq,0}, \mathcal{F}_0)$. $\vec{H}_{\mathcal{H}}$ is the Hessian matrix of the Hamiltonian, which is the Jacobian of the gradient of the Hamiltonian. 

Besides the tangent direction, the predictor algorithm also needs to specify the steplength of the vector. The curve $(\vec{q}_{eq}(s),\mathcal{F}(s))$ can be parametrized by a parameter $s$, such that the steplength is given by $|(\partial \vec{q}_{eq}/\partial s,\partial \mathcal{F}/\partial s)| \times \Delta s$. If parameter continuation is used, the parameter $s$ is proportional to the external parameter such that $s=\alpha \mathcal{F}$, this however leads to problems at folds for which the steplength becomes infinite because $|\partial \vec{q}_{eq}/\partial \mathcal{F}|=\infty$. In the commonly used pseudo-arclength continuation algorithm, the parameter $s$ is parametrized by the length along the curve $s=\int_0^s \alpha |(\partial \vec{q}_{eq}/\partial s,\partial \mathcal{F}/\partial s)| {\rm d}s$  such that the steplength is always finite. 

Note that in order to use path following methods to obtain equilibrium solutions, it is required to start from an initially known equilibrium solution. It is therefore essential that an initial solution can be found, this can sometimes be difficult since there is no general method to find these initial solutions.

\section{Example of path following method}
\label{switchexample}
\begin{figure}[]
\sidecaption
\includegraphics[scale=0.5]{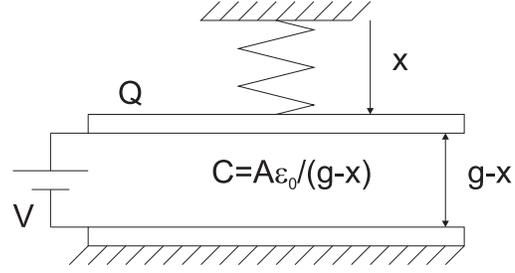}
\caption{An electrostatically actuated capacitor of which one plate is connected to a spring.}
\label{CapMEMS}       
\end{figure}

As an example of the methods discussed in sections \ref{enwork}-\ref{pathfollowing} we consider the problem of finding the equilibra of a voltage controlled capacitor connected to a spring from its energy and work equations. A schematic picture of this system is shown in figure \ref{CapMEMS}. 

The system has two generalized coordinates $x$ and $Q$. The internal energy and work of the system can be used to derive the Hamiltonian:
\begin{eqnarray}
\mathcal{H}&=& U_{strain}+U_{el}-W_{el} = \frac{1}{2}k x^2+\frac{Q^2 (g-x)}{2 A \epsilon_0}-\int_0^Q {V_{ext,eq}(Q) dQ}
\end{eqnarray}
Note that for the conservation of energy to be valid, it is necessary that the externally applied voltage $V_{ext,eq}(Q)$ is taken such that the system is always in equilibrium along the path, otherwise the kinetic energy of the system would become non-zero. The generalized force vector $\vec{F}$ is found from the gradient of the Hamiltonian:
\begin{eqnarray}
\vec{F}=-\nabla \mathcal{H}&=& -\left( \frac{\partial \mathcal{H}}{\partial x},\frac{\partial \mathcal{H}}{\partial Q} \right) \nonumber \\
\label{Fex}
&=&\left( -k x+\frac{Q^2}{2 A \epsilon_0}, V_{ext}(Q)-\frac{Q (g-x)}{A \epsilon_0} \right) =0
\end{eqnarray}
The Hessian matrix of the Hamiltonian is found\footnote{The method presented here is slightly different from the method which we presented in \cite{Stulemeijer}. In \cite{Stulemeijer} the electrical generalized force equation $V_{ext}=Q/C$ was eliminated before taking the Jacobian of the force vector. In this derivation we keep all generalized force equations to have a square Hessian matrix.} to be:

\begin{eqnarray}
-\vec{H}_{\mathcal{H}} &=& \vec{J}_{\vec{F}} = -\left( 
\begin{array}{cccc}
\frac{\partial^2 \mathcal{H}}{\partial x^2} &,& \frac{\partial^2 \mathcal{H}}{\partial x \partial Q} \\
\frac{\partial^2 \mathcal{H}}{\partial Q \partial x} &,& \frac{\partial^2 \mathcal{H}}{\partial Q^2}
\end{array} \right) = \left(
\begin{array}{cccc}
-k &,&\frac{Q}{A \epsilon_0} \\
\frac{Q}{A \epsilon_0} &,&  \frac{\partial V_{ext,eq}}{\partial Q}-\frac{(g-x)}{A \epsilon_0}
\end{array} \right) 
\end{eqnarray}
\begin{figure}[]
\sidecaption
\includegraphics[scale=0.65]{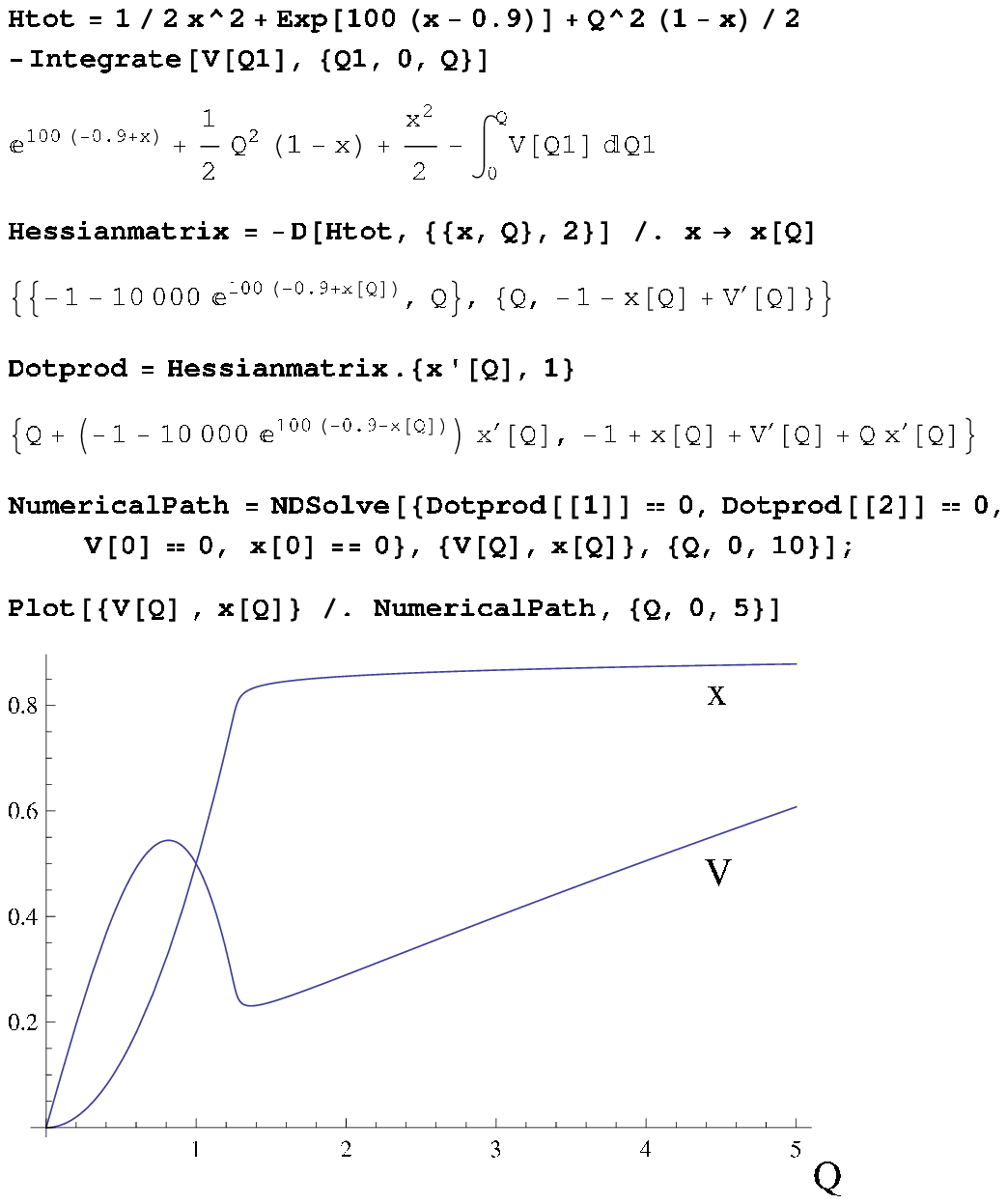}
\caption{This script in Mathematica  shows how the $V_{eq}-Q$ curve of an electrostatically actuated MEMS switch with a non-linear exponential spring contact can be obtained using path following as described in section \ref{switchexample}. The spring energy is given by $\frac{1}{2}k x^2+\exp{(c_1(x-g+\delta))}$, the electrical energy is $Q^2/2C$ and the capacitance $C=A \epsilon_0/(g-x)$, with $k=g=A \epsilon_0=1$,$c_1=100$ and $\delta=0.1$. The path is followed starting from the solution $x=V=Q=0$. The $V_{eq}-Q$ curve is identical to that in figure \ref{Energy}. The $x-Q$ curve is also plotted.}
\label{Examplecalc}       
\end{figure}

Thus equation (\ref{Hessian}) gives us:
\begin{eqnarray}
\label{Hmatcap}
-\vec{H}_{\mathcal{H}} (dx,dQ)= \left( -k dx + \frac{Q}{A \epsilon_0} dQ, \frac{Q}{A \epsilon_0} dx  + \frac{\partial V_{ext}}{\partial Q} dQ- \frac{(g-x)}{A \epsilon_0}  dQ \right) = 0
\end{eqnarray}
Dividing these two equations by $dQ$ and rearranging shows that:
\begin{eqnarray}
\label{path1}
\frac{{\rm d} x}{{\rm d}Q}&=&\frac{Q}{k A \epsilon_0} \\
\label{path2}
\frac{\partial V_{ext,eq}}{\partial Q} &=& \frac{{\rm d} V_{ext,eq}}{{\rm d} Q}=\frac{1}{C}-\frac{Q^2}{k A^2\epsilon^2_0}
\end{eqnarray}

At an equilibrium point at coordinate $(x_{eq}, Q_{eq},V_{ext,eq})$ equations (\ref{path1}) and (\ref{path2}) provide the tangent vector to the solution curve $(\frac{dx}{dQ} dQ, dQ,\frac{d V_{ext,eq}}{d Q} dQ)$. In figure \ref{Examplecalc} we show how the equations in this section can be implemented in a Mathematica \cite{mathematica} script to determine the $V-Q$ curve of an electrostatically actuated MEMS switch with a non-linear exponential spring contact. In this example we have used the NDSolve function in Mathematica to follow the solution path solely from the tangent vector function, without using a corrector method. The different types of energy in this system are shown in figure \ref{Energy}.

Dedicated interactive numerical path following packages \cite{Govaerts} like \textsc{Matcont} \cite{matcont} and \textsc{Auto} \cite{auto} include corrector methods and stability analysis. They are therefore more robust and convenient to treat complex problems. An example in \textsc{Matcont} is discussed in section \ref{matcontex}.

\section{Stability}
\label{stability}
The system is in a static equilibrium state $\vec{q}_{eq}$ when the sum of all forces is zero and all generalized momenta are zero. This equilibrium is only stable if for all possible infinitesimal displacement vectors $\vec{dq}_{eq}$, the Hamiltonian increases such that $\mathcal{H}(\vec{q}_{eq}+\vec{dq}_{eq})>\mathcal{H}(\vec{q}_{eq})$. Any displacement would therefore violate the conservation of energy and thus the system will remain in state $\vec{q}_{eq}$ forever.

The local Taylor expansion of the Hamiltonian is given by:
\begin{equation}
\label{taylor}
\mathcal{H}(\vec{q}+\vec{dq})=\mathcal{H}(\vec{q})+\vec{\nabla}\mathcal{H}(\vec{q}) \cdot \vec{dq}+\frac{1}{2} \vec{dq} \cdot \vec{H}_{\mathcal{H}}(\vec{q}) \vec{dq}
\end{equation}

At a static equilibrium point, the gradient of $\mathcal{H}$ is zero according to equation (\ref{Feq}), so the system is stable if the right term of equation (\ref{taylor}) is positive for all $\vec{dq}$. The vector $\vec{dq}$ can be written as $\vec{dq}=\sum_i^N a_{dq,i} \vec{\hat{q}}_{H,i}$, where $\vec{\hat{q}}_{H,i}$ are the eigenvectors of the Hessian matrix with corresponding eigenvalues $\lambda_i$. Therefore in equation (\ref{taylor}) the term $\frac{1}{2} \vec{dq} \cdot \vec{H}_{\mathcal{H}}(\vec{q}) \vec{dq}=\frac{1}{2} \sum_i^N  \lambda_i a^2_{dq,i} \vec{\hat{q}}^2_{H,i}$, is always positive if all eigenvalues $\lambda_i$ of the Hessian matrix are positive. Points at which one eigenvalue becomes zero are called folds. Points at which more than one eigenvalue become zero are called bifurcations. Since these zero crossings cause a sign change of the eigenvalues, the stability of the system usually changes at these points.

\begin{figure}[]
\begin{minipage}[b]{0.5\linewidth}
\centering
\includegraphics[scale=0.32]{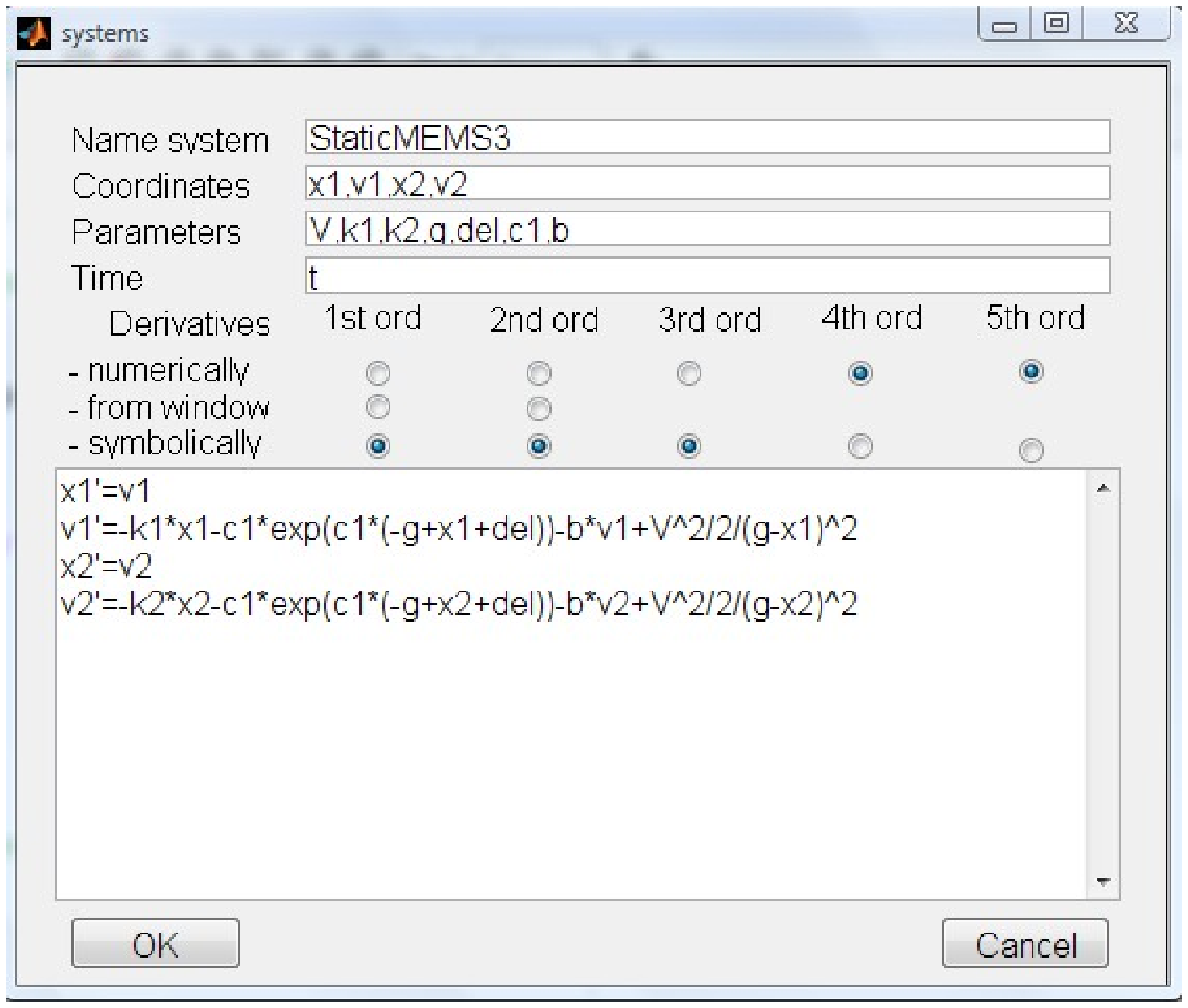}
\end{minipage}
\begin{minipage}[b]{0.5\linewidth}
\centering
\includegraphics[scale=0.4]{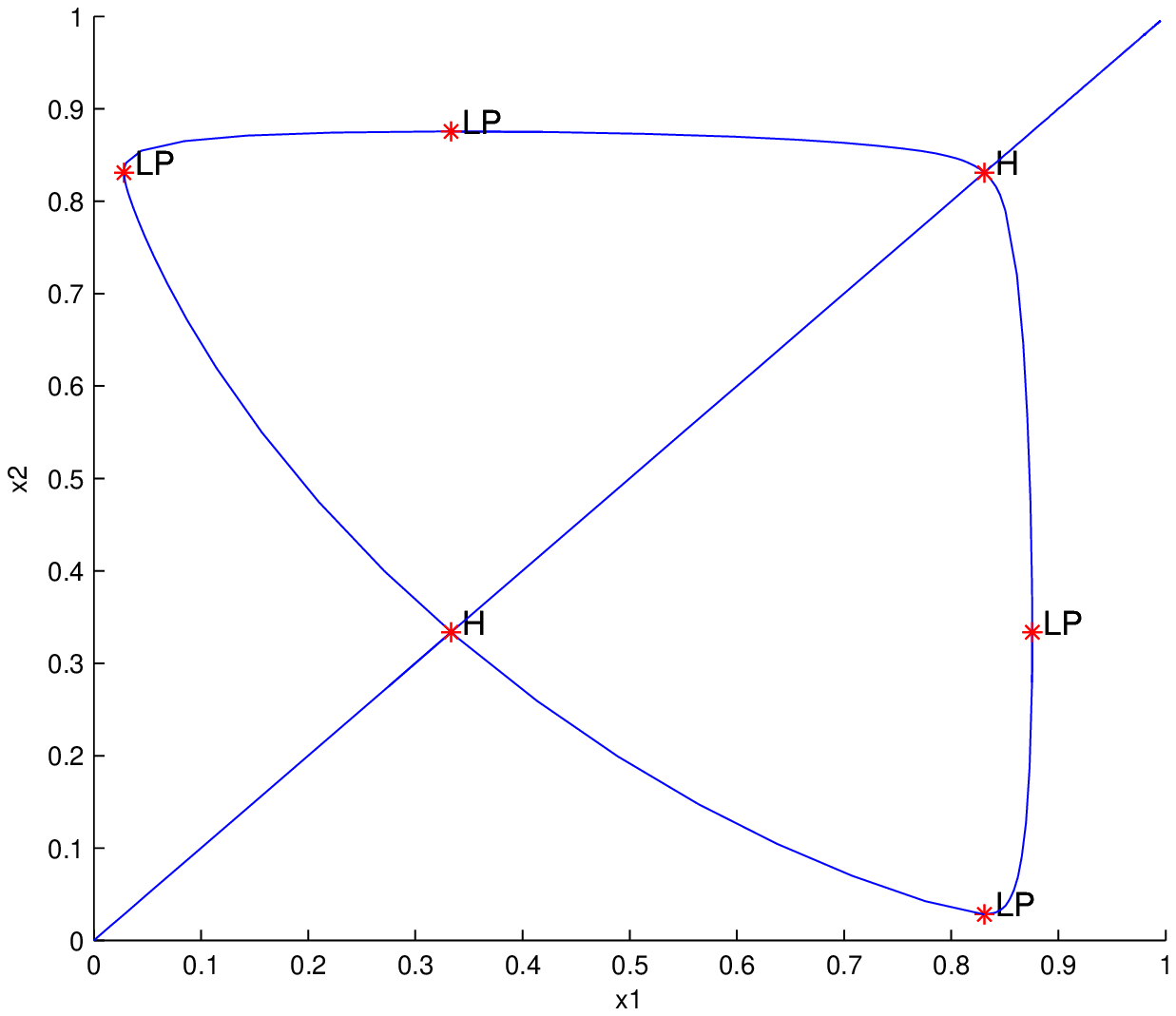}
\label{fig:figure2}
\end{minipage}
\caption{Left: Model equations in \textsc{Matcont} for two identical switches with the same parameters as the switch in figure \ref{Examplecalc}. Right: Calculated equilibrium curves. Folds (LP) and bifurcations (H) are automatically detected by the software. The folds (LP) correspond to the pull-in and release instabilities of one switch, at bifurcations (H) both switches are unstable for pull-in or release.}
\label{StaticMEMS}       
\end{figure}


\section{Numerical path following in \textsc{Matcont}}
\label{matcontex}
As an example of the use of \textsc{Matcont} for path following, we take two identical uncoupled switches with the same parameters as in figure \ref{Examplecalc} and enter the force and momentum equations (\ref{Hamilton}) in the system window shown on the left side of figure \ref{StaticMEMS}. To improve the convergence of the calculation a mass and damping term have been added to the force equations. On the right side of the figure the calculated path following curves are shown.
Path following programs like \textsc{Matcont} monitor the eigenvalues of the Hessian on the equilibrium path to determine the stability of the system and plots the folds (LP) and bifurcation points (H) on the graph (see figure \ref{StaticMEMS}). As an example two identical capacitive MEMS switches are simulated in in figure \ref{StaticMEMS}. All eigenvalues of the Hessian are positive on the line from (0,0)-H. At the bifurcation point H $(\frac{1}{3},\frac{1}{3})$, two eigenvalues become zero and both switches become unstable towards pull-in. On the paths H-LP only one switch is unstable and the other is open (one negative eigenvalue), on the path H-H both switches are unstable (two negative eigenvalues). On the segments LP-LP between the release and pull-in point of one of the switches and on the segment H-(1,1), where both switches are closed, the system is stable and all eigenvalues are positive.

\section{Finite element method path following simulations}
\label{fem}
The examples given up to now have a low number of degrees of freedom, such that their energy equations can be entered manually. To model continuous systems, for which the energy and work equations are given by the volume integrals in equations (\ref{ukin}-\ref{wstrain}), dedicated finite element method (FEM) software is available. These packages mesh the volume and construct partial differential equations that provide the equations of motion for all degrees of freedom. Moreover they provide efficient solvers, such that systems with thousands or even millions of degrees of freedom can be solved.

\begin{figure}[]
\vspace{0.5cm}
\includegraphics[scale=0.45]{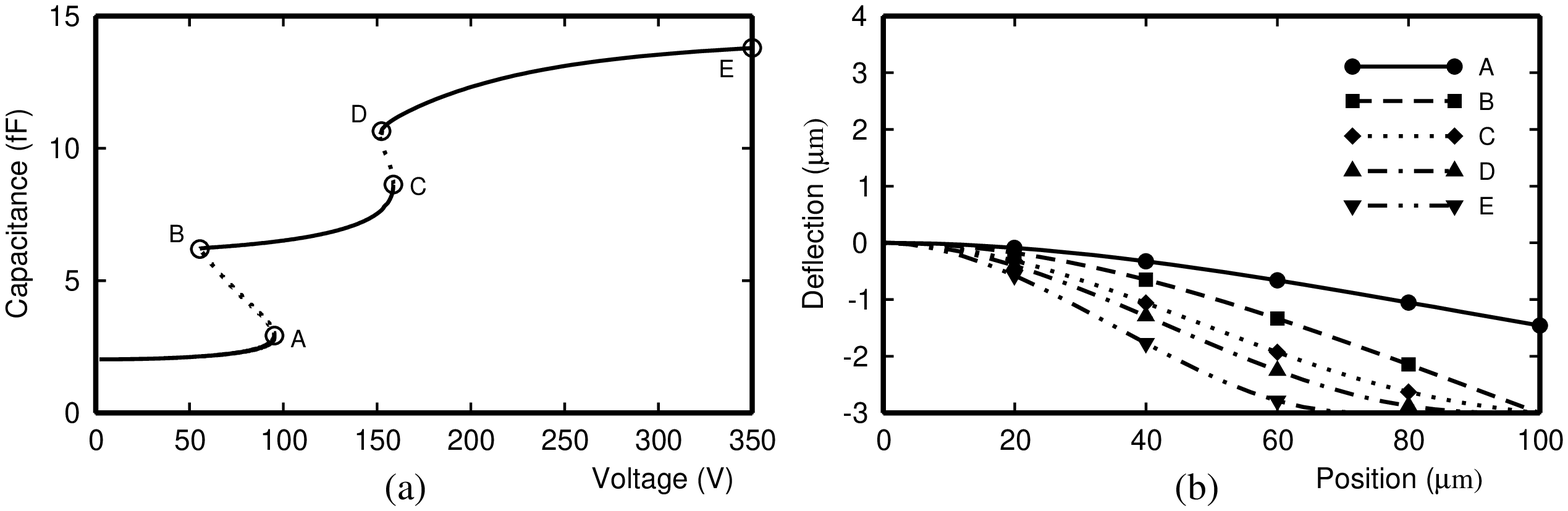}
\caption{Simulation of the equilibria of a voltage actuated single clamped Euler beam (adapted from \cite{Stulemeijer}). a) The capacitance voltage curve shows two pull-in (A,C) and two release (B,D) instabilities. The pull-in voltage at point A corresponds to the transition from no contact (float) to contact with a single point (pinned). The pull-in at point C corresponds to the transition from single point contact (pinned) to contact over a finite length (flat). The stable equilibria are indicated by solid lines and the unstable by dotted lines. b) The displacement shape of the beam at 5 points (A-E) on the $C-V$ curve.}
\label{FEM}       
\end{figure}

Some finite element packages do provide basic path following methods to solve non-linear problems, however the control over the followed path and the detection of stability is less advanced than in the dedicated path following packages \cite{Govaerts} discussed in the previous section. To analyze non-linear MEMS systems with path following methods in a FEM package, we have written a script for \textsc{Comsol} \cite{comsol}. The script is similar to that presented in \cite{moller} and predicts the initial condition for the next calculation and uses the solver of \textsc{Comsol} as a corrector. The prediction direction is a simple linear extrapolation based on the two previous solutions. The predicted steplength is based on the previous steplength and the curvature of the path. The obtained corrected solution is discarded if its direction or distance is too far from the predicted solution. Since \textsc{Comsol} provides direct access to the stiffness matrix (which is the Hessian of the Hamiltonian), the stability of the solutions can be determined by checking the sign of the eigenvalues of this matrix. By making use of the stiffness matrix, more sophisticated predictor algorithm scripts can be developed.

In figure \ref{FEM} an example of a FEM simulation of a voltage actuated single clamped beam is shown, which was obtained using the \textsc{Comsol} script. The results correspond well to the analytical results in \cite{Gorthi}. More details and examples of FEM numerical path following can be found in \cite{Stulemeijer}. Although this example is still relatively simple, the FEM path following method can theoretically be applied to analyze any non-linear system of which the Hamiltonian $\mathcal{H}$ can be written in terms of the analytic volume integrals (\ref{ukin}-\ref{wstrain}). 

\section{Conclusions}

It has been shown that path following can be a powerful mathematical tool to determine the equilibrium solutions of electromechanical systems and their stability from the energy and work expressions. The path following of the electromechanical equilibrium solutions can be performed by analytical methods, dedicated numerical software or finite element method software. Although this paper has focused on static systems, the numerical path following technique can also be applied to periodic dynamic systems (see \cite{Doedel}), which even extends its usefulness as mathematical tool for analyzing electromechanical systems.

\end{document}